\documentclass[]{aa}
\usepackage{graphicx}
\usepackage{amsmath}
\usepackage{amsfonts}
\usepackage{amssymb}
\usepackage{graphicx,natbib,color}
\usepackage[british]{babel}
\bibpunct{(}{)}{;}{a}{}{,}
\newcommand{\ignore}[1]{}

\newcommand{\kpc}{\mathrm{kpc}}

\newcommand{\ccm}{\mathrm{cm}^{-3}}

\newcommand{\replace}[2]{{ #2}}
\begin{document}
   \title{Galactic Wind Shells and High Redshift Radio Galaxies}
   \subtitle{On the Nature of Associated Absorbers}

   \author{Martin Krause
           }

   \offprints{Martin Krause}

   \institute{Astrophysics Group, Cavendish Laboratory, Cambridge CB3 0HE,
	United Kingdom \\\email{M.Krause@lsw.uni-heidelberg.de}
         }

   \date{Received August 19, 2004; accepted date}

   \abstract{A jet is simulated on the background of a galactic wind 
	headed by a radiative bow shock.
        The wind shell, which is due to the radiative bow shock, 
	is effectively destroyed by the impact of the jet cocoon,
        thanks to Rayleigh-Taylor instabilities.
	Associated strong HI absorption, and possibly also molecular emission, 
       	in high redshift radio galaxies
        which is observed preferentially in the smaller 
        ones may be explained 
        by that model, which is an improvement of an earlier 
	radiative bow shock model.
        The model requires temperatures of $\approx 10^6$~K in the 
	proto-clusters hosting these objects, and may be tested by 
        high resolution spectroscopy of the Ly$\alpha$ line.
	The simulations show that -- before destruction -- the jet cocoon 
	fills the wind shell entirely for a considerable time with 
	intact absorption system. 
	Therefore, radio imaging of sources smaller than the critical size
	should reveal the round central bubbles, if the model is correct.
   }

   \maketitle
%

\section{Introduction}\label{intro}
Radio galaxies at high redshift ($z\gtrsim2$) show huge (similar to the radio size), 
luminous ($\approx 10^{43}-10^{45}$~erg/s) Ly$\alpha$
halos \citep{Rottea96,vOea97,Reulea03}. Being the progenitors of today's brightest cluster galaxies \citep{Cea01},
they pinpoint proto-clusters of galaxies \citep{Venea02,Venea03}. 
Many of the systems smaller than $\approx 50$~kpc show associated absorption, preferentially 
on the blue wing of the emission line \citep{vOea97,Binea00,dBrea00}. While the low column density absorbers 
($N_\mathrm{HI} \approx 10^{13} - 10^{14}$~cm$^{-2}$) are probably mainly due to the Ly$\alpha$ forest,
the high column density ones ($N_\mathrm{HI} \approx 10^{18} - 10^{20}$~cm$^{-2}$) are found much more frequently
than expected from the Ly$\alpha$ forest and hence belong to the radio galaxy \citep{vOea97,WJR04}. High resolution spectroscopy 
has revealed that some of the latter are made up of overlapping smaller ones. But some others have been confirmed 
\citep{WJR03,WJR04}. 

The frequency of this phenomenon points to common circumstances in the surroundings of these most massive 
objects in the young universe. A sound understanding of the emission line structure therefore promises insight into
their assembly. 
Here, I present \replace{a hydrodynamic simulation}{hydrodynamic simulations} 
of a jet born inside a galactic wind shell, the latter having a radiative 
bow shock. This can be regarded as a modification of an earlier model \citep{mypap02a}, where the jet itself 
produced the radiative bow shock. The big absorption systems might be identified with such a geometrically thin, 
dense shell, as I will discuss in the following.

\ignore{The letter is organised as follows:} 
Section~\ref{review} reviews the available models from the literature.
Section~\ref{jpw} introduces 
the new one. The computation is presented in section~\ref{sim}, 
and the implications and observational tests are discussed in section~\ref{discu}.

\section{Review of absorber models}\label{review}
\begin{figure*}
\centering
\includegraphics[width=0.99\textwidth]{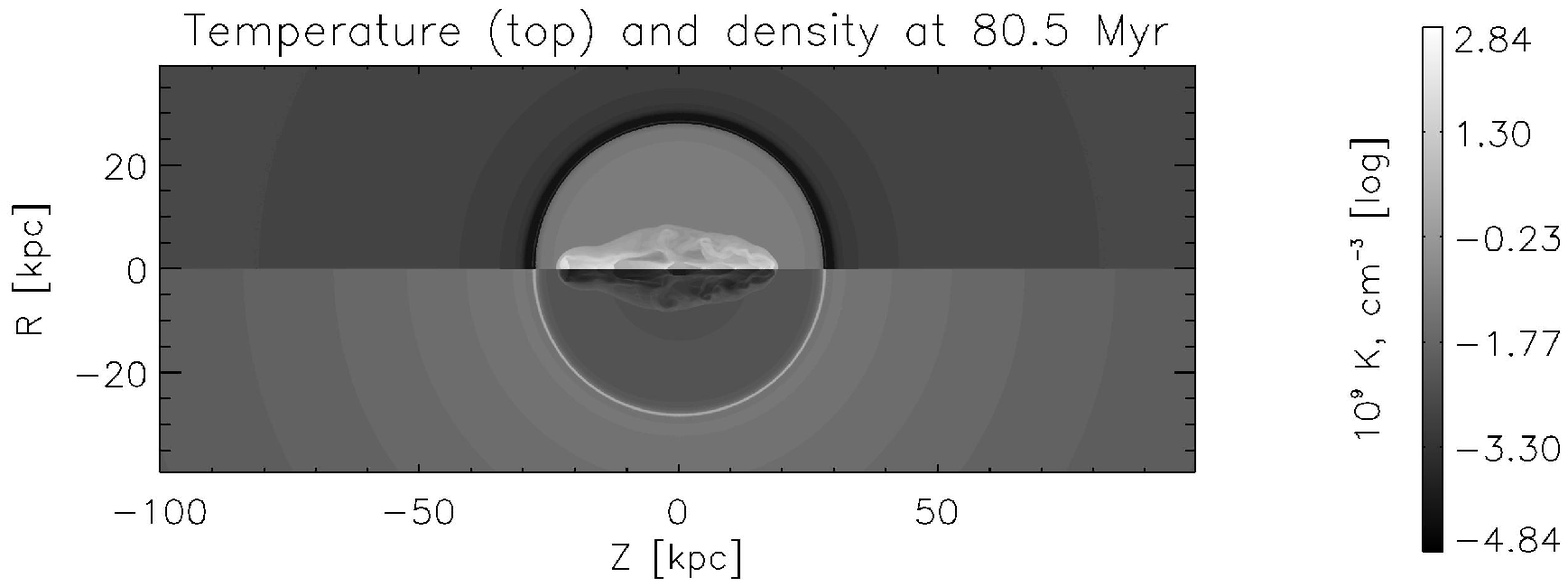}
\includegraphics[width=0.99\textwidth]{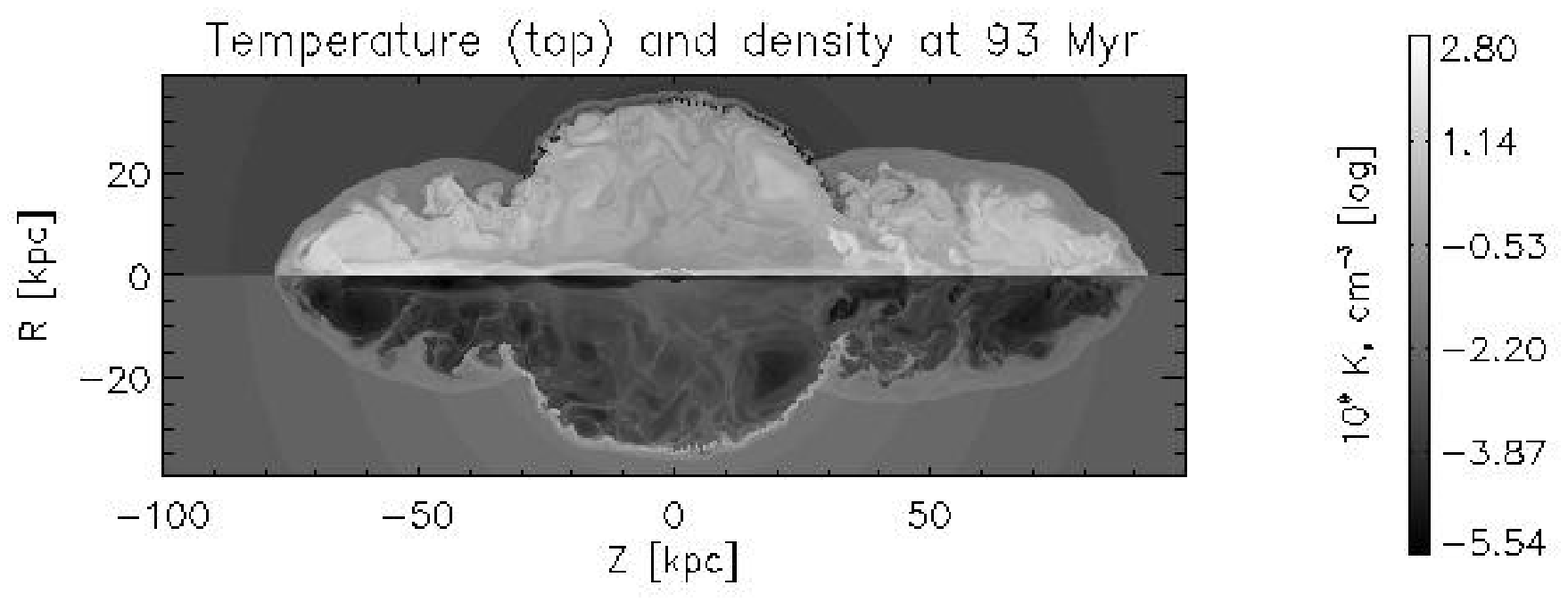}
\caption{\small Density and temperature distribution after 80.5~Myr 
	(0.5~Myr after jet start), 93~Myr (13~Myr after jet start) 
	for the low resolution simulation. \label{dtl}}
\end{figure*}
\begin{figure*}
\centering
\includegraphics[width=0.99\textwidth]{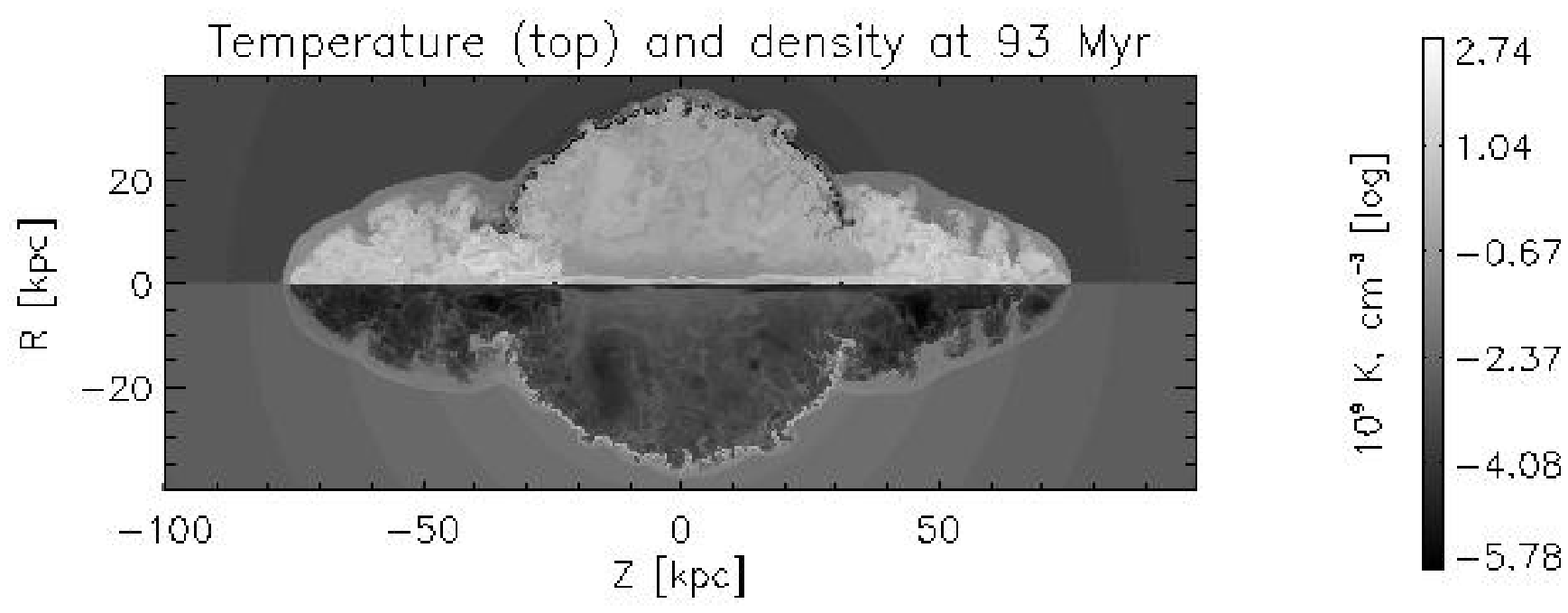}
\includegraphics[width=0.99\textwidth]{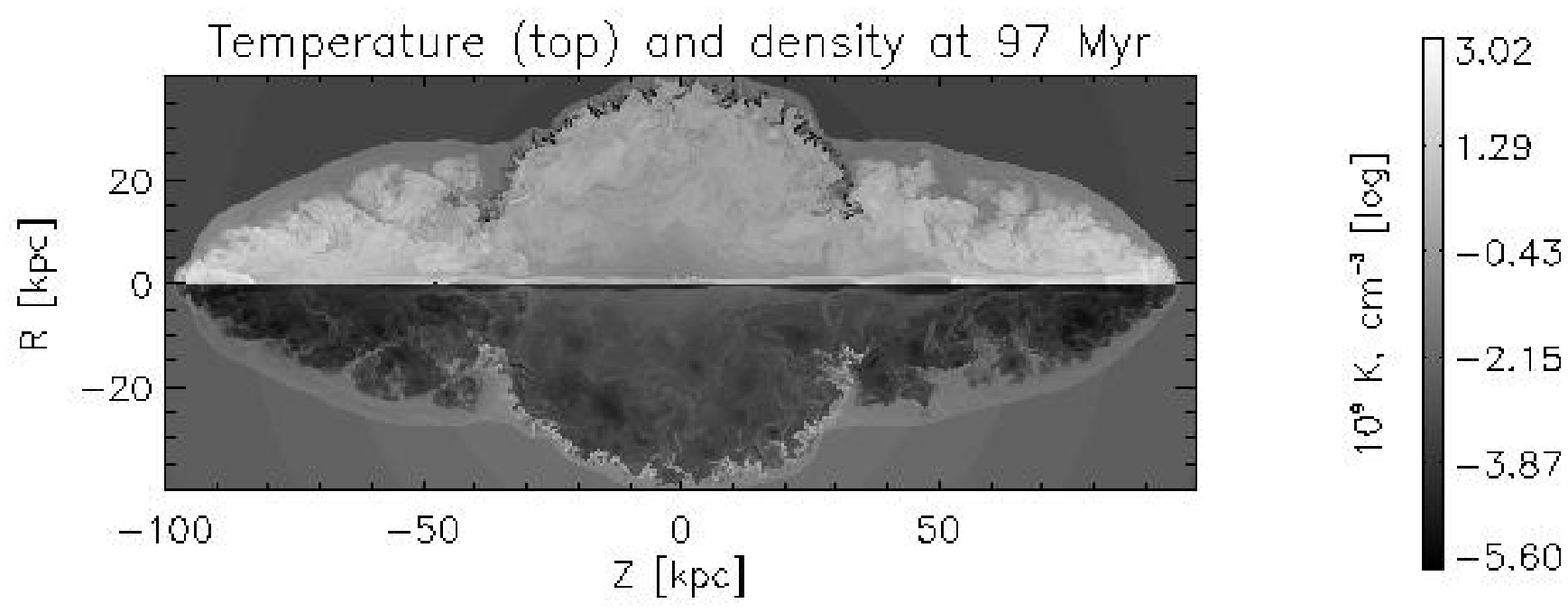}
\caption{\small Density and temperature distribution after 93~Myr 
	(13~Myr after jet start), 97~Myr (17~Myr after jet start) 
	for the high resolution simulation. Only part of the grid
	is shown in the radial direction. Comparison to Fig.~\ref{dtl}
	shows the same overall evolution. Details that differ include
	the more evident asymmetry in the low resolution simulation,
	and the slower growth of the outer jet bubbles in the high
	resolution case.  \label{dth}}
\end{figure*}

\subsection{$\it 10^{12}$clouds in an ionisation cone}
In this model, proposed by \citet{vOea97}, $\approx 10^{12}$ clouds with a density of $n_\mathrm{c}\approx 100$~cm$^{-3}$,
about $10^8 M_\odot$ in total,
are distributed throughout the region occupied by the radio galaxy.  A central quasar, hidden from the observer by a central gas and dust torus,
ionises the clouds within two cones, where the gas is also stirred up by interaction with the jet 
(observations indicate a typical line width of 1000~km/s). Outside the cone,
the clouds are neither ionised by the quasar nor stirred up by the radio source, 
and hence are able to absorb the Ly$\alpha$ emission of the
ionised clouds in a narrower velocity range (typically $10-100$~km/s). 
The smaller sources are located in denser environments, which also produce more H{\sc i} absorption.

This model cannot explain the tendency of these absorption systems to be 
blueshifted ($\approx 100-200$~km/s). 
\replace{
At least in one case the absorber has been detected in C{\sc iv} absorption
\citep{Binea00}, and a lower metallicity compared with the one of 
the emitting gas has been infered. In this case, the absorbing 
gas has to be physically different from the emitting one.}{
In at least one case the absorber has been detected in CIV \citep{Binea00}, 
who inferred a lower metallicity in the absorber than in the emitting plasma. 
This would require that the absorbing gas is physically distinct from the 
line emitting plasma.}

\subsection{Extended low density shell}
\citet{Binea00} also infered a high ionisation of the absorbing gas. The lack of an obvious source of the ionising photons 
forced them to conclude that the shell was a low density quiescent absorbing screen surrounding the radio galaxy.
They favoured a density of $10^{-3}-10^{-2}$~cm$^{-3}$, because such an absorber would be ionised by the metagalactic background
radiation.
They proposed an evolutionary scenario: Once the radio source reaches the screen, the increased pressure compresses the denser
parts, that from now on can only be seen when inside the ionisation cone, while the rest gets completely ionised. 

This scenario also has problems with the velocity structure. Suppose there would be a quiescent gas halo at a temperature of a few times 
$T=10^4$~K and a density of $n=10^{-3}$~cm$^{-3}$ surrounding the centre of the radio galaxy in a distance of $\approx 50$~kpc. 
The cooling time would be $t_\mathrm{c} = k_\mathrm{B} T/n \Lambda \approx$~Myr, where 
$\Lambda\approx 10^{-22}$~erg~cm$^3/\mathrm{s}$ is the cooling function \citep[e.g.][]{SD93}. 
After that time, which is much shorter than the expected age of the 
observed radio sources, a cooling flow should be established, and the absorbers should appear on the red wing, not on the blue one, 
as observed. It was concluded that the absorbing gas has to be outflowing due to a galactic wind \citep{WJR03}.
In this case, the density distribution would be expected to be smooth ($\approx r^{-2}$). When the source expands,
it should gradually reduce the neutral column. This has not been shown to be able to accommodate the observed
bimodal distribution of column densities.
Another problem arises from the temperature in such an outflow, which is given by
\begin{eqnarray*}
T  &=&  
\frac{(\gamma-1) Lt}{(4/3) \pi r^3 n k_\mathrm{B}} 
     =  10^7 \,\mathrm{K} \times \\ 
& & \frac{L}{10^{42}\,\mathrm{erg/s}}
 \frac{t}{100\,\mathrm{Myr}} \frac{10^{-3}\,\mathrm{cm}^{-3}}{n}
\left(\frac{25\,\mathrm{kpc}}{r}\right)^{3}   \enspace,
\end{eqnarray*}
where $\gamma$ is the adiabatic index, $L$ is the driving power of the galactic wind, a few supernovae per century, 
$r$ is the radius of the shell,
and $t$ the time for which the wind is active. So, one would expect the gas in these outflows to be too hot to contain
neutral hydrogen. If they should contain denser condensations, that provide the absorption, the ionisation problem reappears.
If the parameters of these winds would be tuned to the right temperature, the cooling flow problem reappears. The gas would be expected 
to loose its pressure support and switch from outflow to inflow.
\begin{figure*}
\centering
\includegraphics[width=0.56\textwidth]{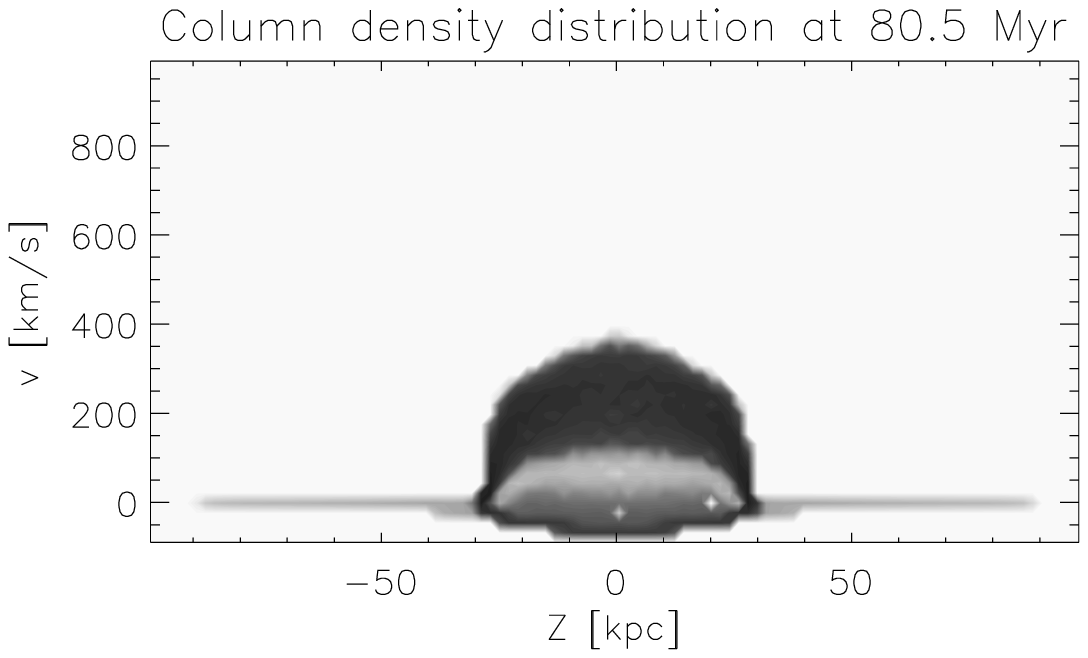}\includegraphics[width=0.42\textwidth]{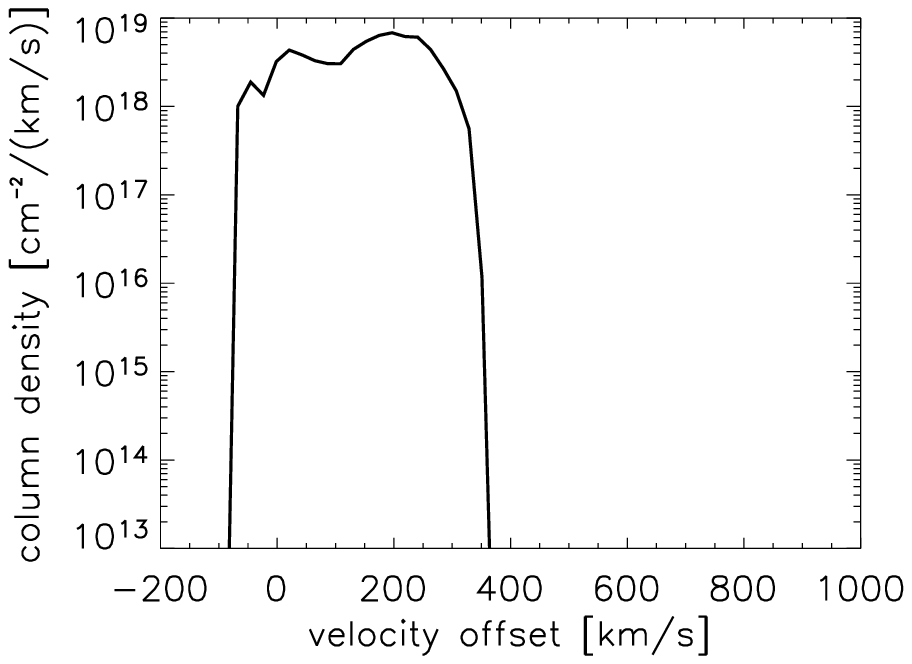}
\includegraphics[width=0.56\textwidth]{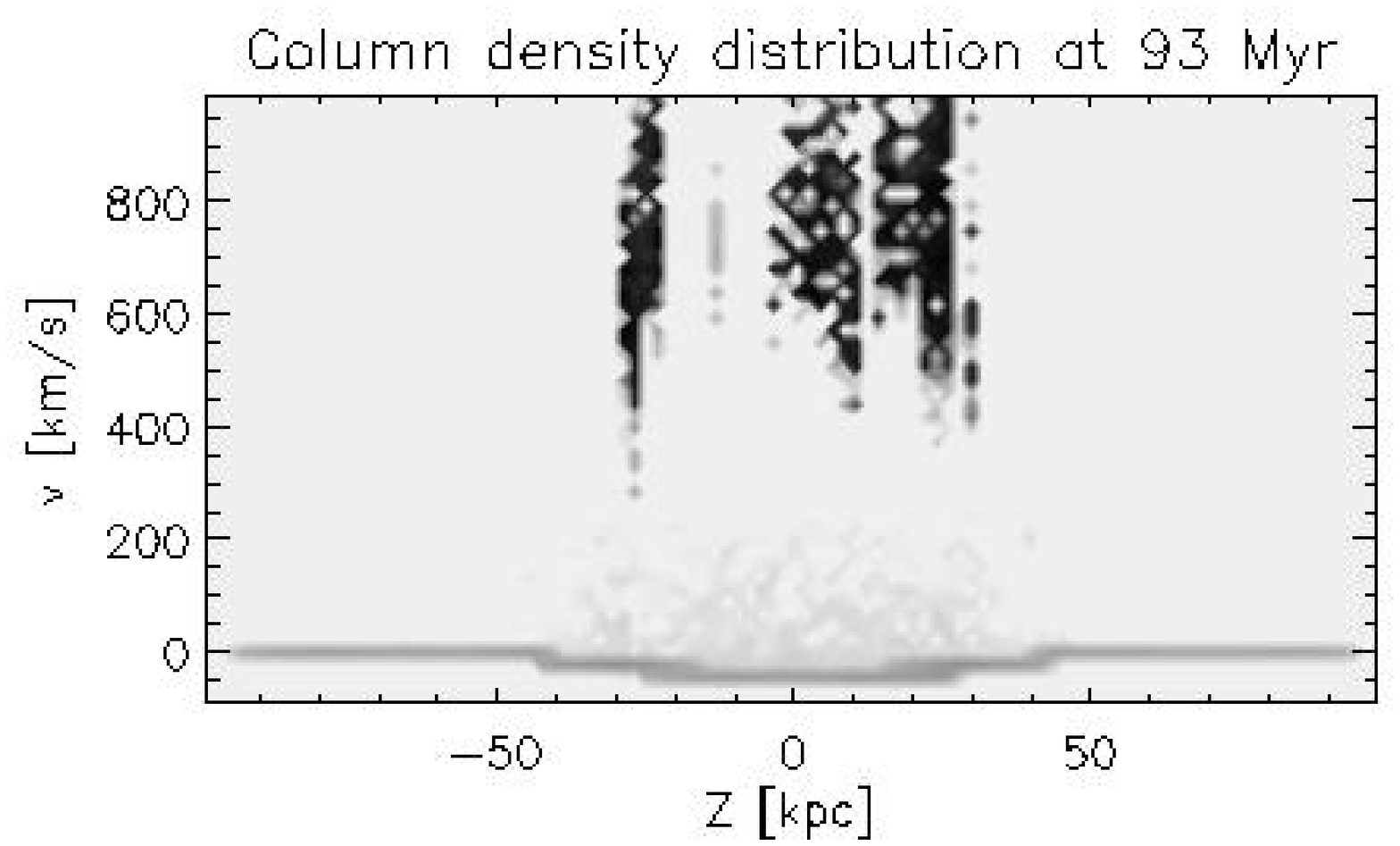}\includegraphics[width=0.42\textwidth]{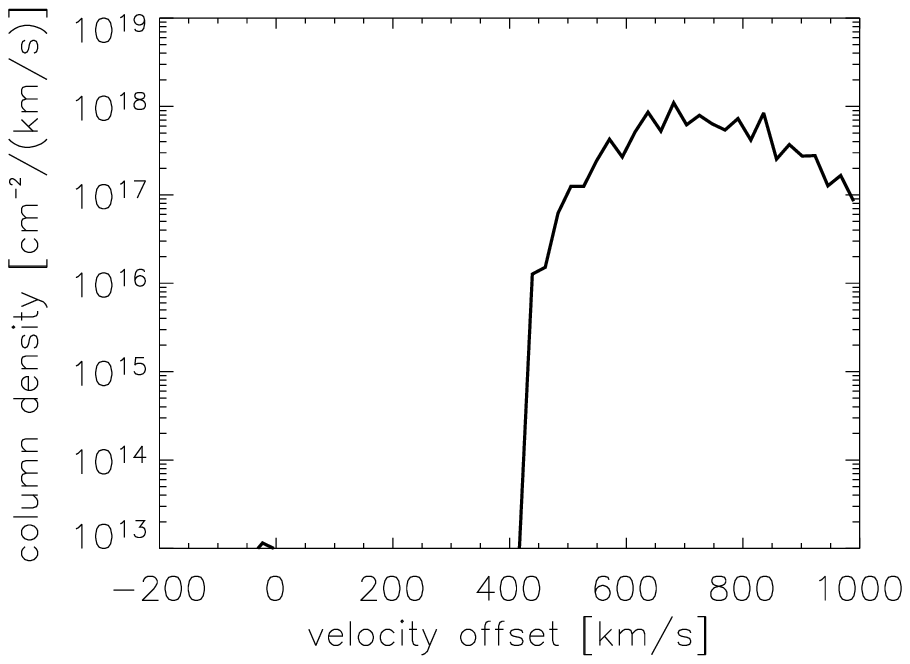}
\caption{\small Column density of neutral hydrogen over Z-position and radial
  velocity (left), and integrated over space (right), after 80.5~Myr 
(top, 0.5~Myr after jet start) and 93~Myr (bottom) 
for the low resolution simulation.
\label{cd1a}}
\end{figure*}
\begin{figure*}
\centering
\includegraphics[width=0.56\textwidth]{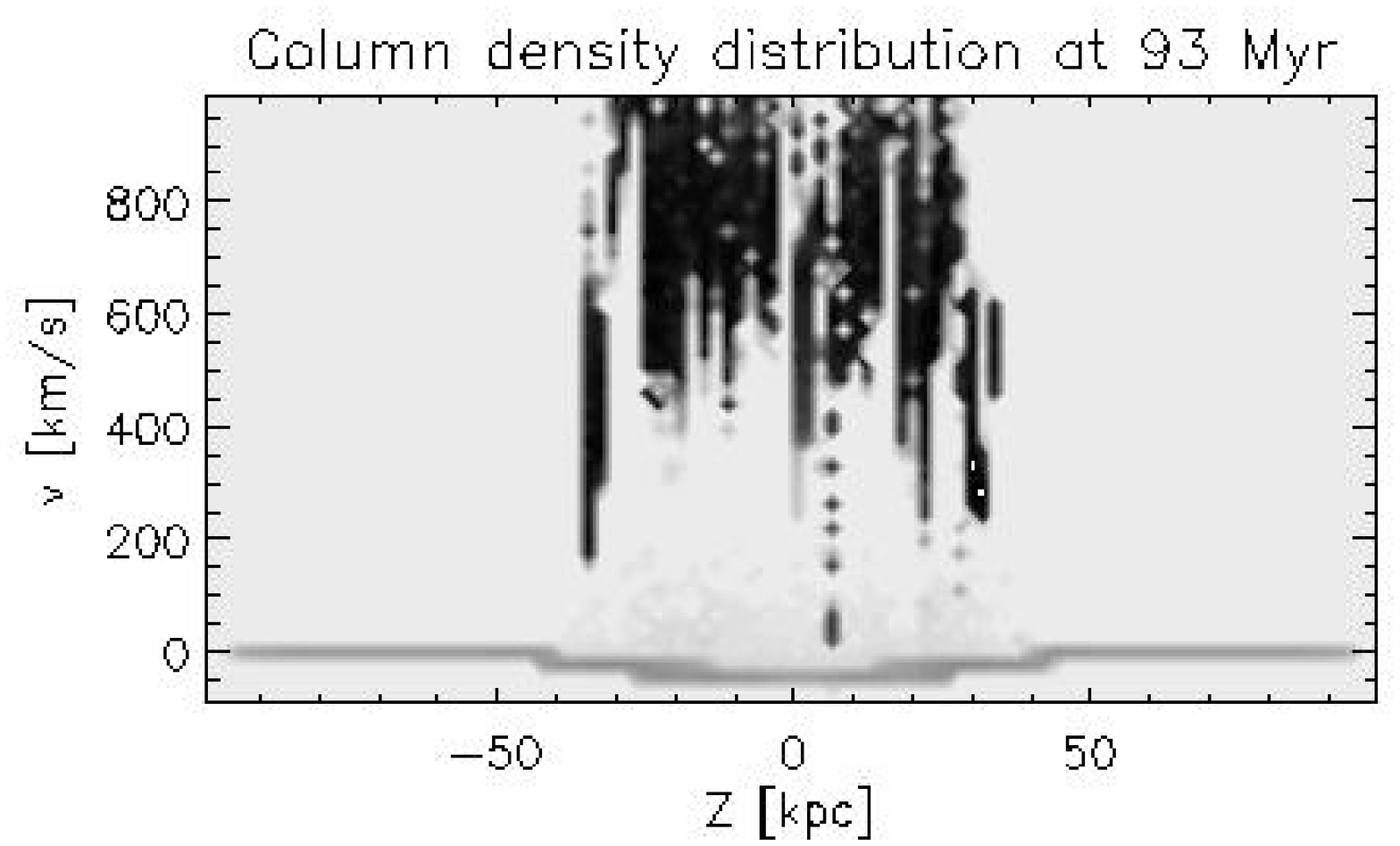}\includegraphics[width=0.42\textwidth]{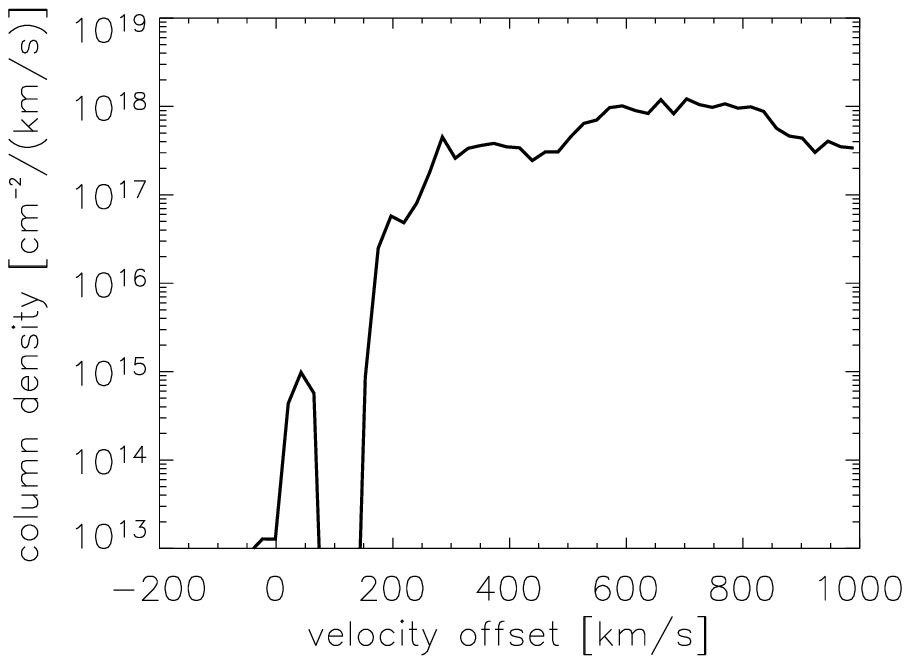}
\includegraphics[width=0.56\textwidth]{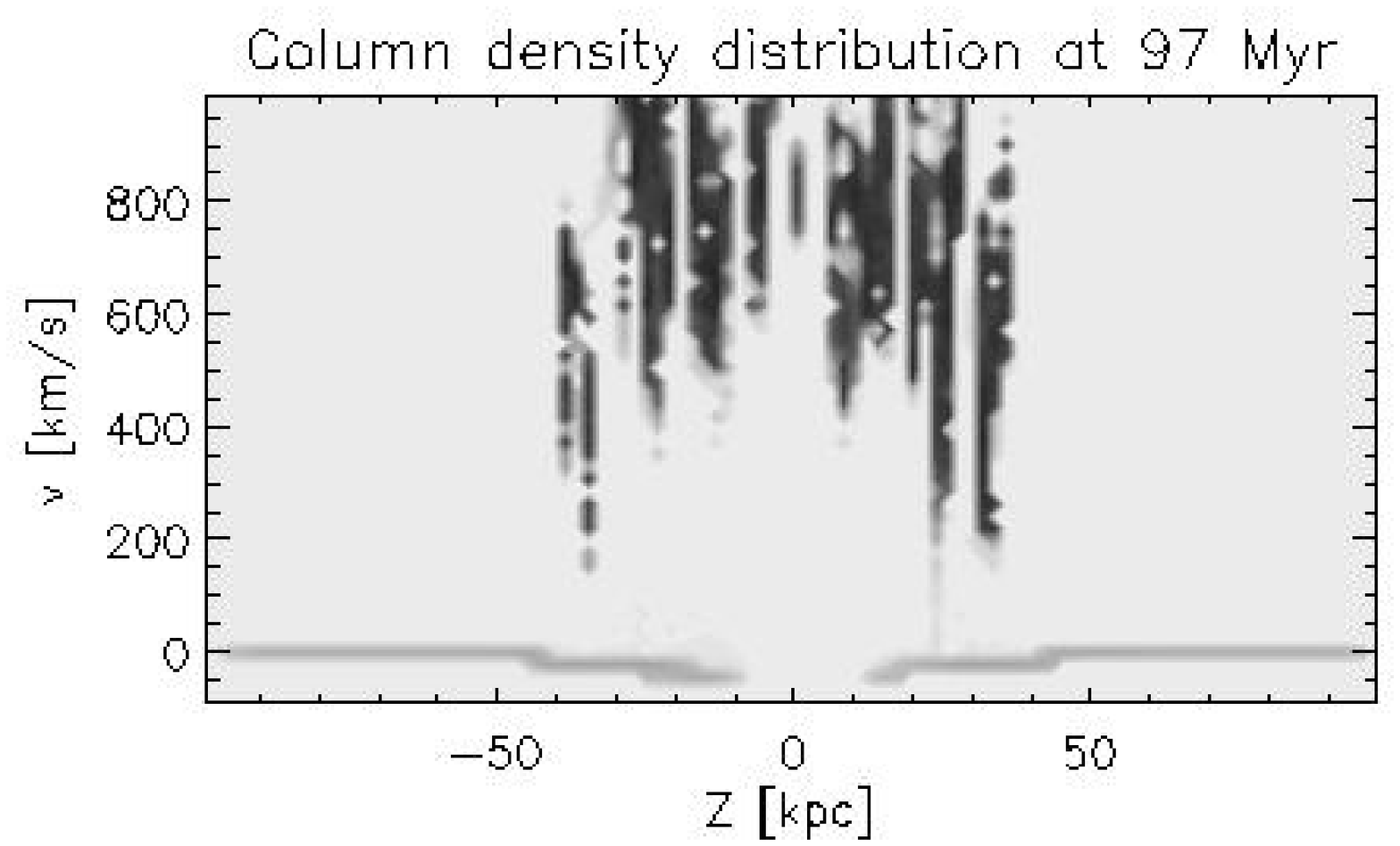}\includegraphics[width=0.42\textwidth]{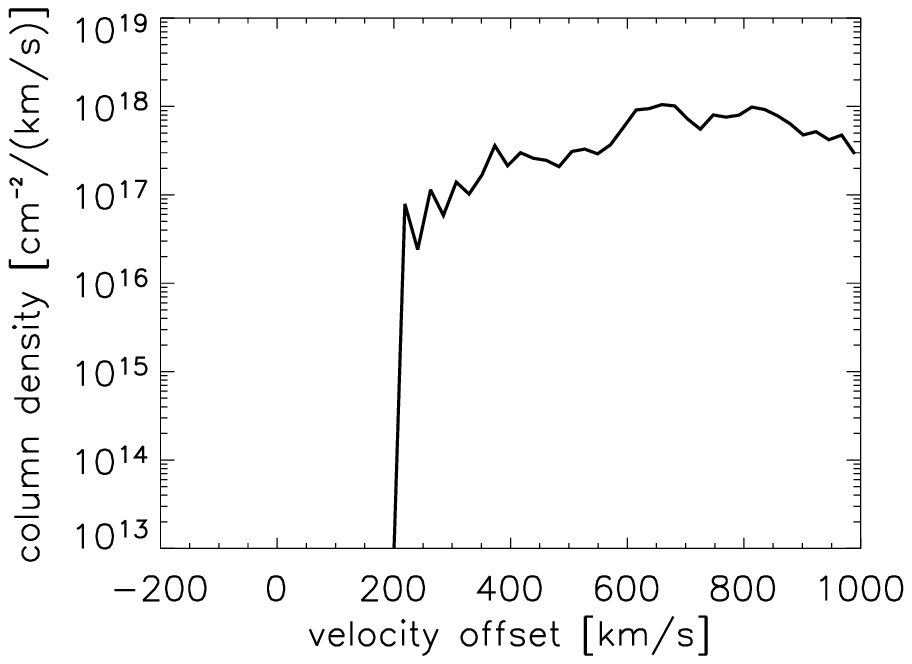}
\caption{\small Column density of neutral hydrogen over Z-position and radial
  velocity (left), and integrated over space (right), after 93~Myr 
(top, 13~Myr after jet start) and 97~Myr (bottom) 
for the high resolution simulation.
\label{cd1b}}
\end{figure*}

\subsection{Jet with radiative bow shock}
Three consequences arise if a jet propagates into a dense medium. First, the bow shock slows down. Second, it may become radiative,
and third, very light jet physics applies \citep{mypap02a,mypap03a}. 
In this case, the bow shock is first spherical, following the same law of motion as the
stellar (and galactic) wind bubble, which reads for constant external density $\rho$:
\begin{equation}
\label{rbow}r_\mathrm{sh} = 12 \, \mathrm{kpc} \, \left(\frac{L}{10^{46}\,\mathrm{erg/s}}
        \frac{m_\mathrm{p} \,\mathrm{cm}^{-3}}{\rho} 
        \left(\frac{t}{10\,\mathrm{Myr}}\right)^{3} \right)^{1/5} 
\end{equation}
\begin{equation}
\label{vbow} v_\mathrm{sh} = 450 \, \mathrm{km/s} \,  \left(\frac{L}{10^{46}\,\mathrm{erg/s}}
        \frac{m_\mathrm{p}\, \mathrm{cm}^{-3}}{\rho} \left(\frac{25\,\mathrm{kpc}}{r_\mathrm{sh}}\right)^{2} \right)^{1/3}
\end{equation}
An external density of a few $m_\mathrm{p}\, \mathrm{cm}^{-3}$ would slow the bow shock velocities down to the observed values.
Such a slow bow shock would heat a surrounding X-ray atmosphere only marginally.
The cooling time due to bremsstrahlung is $t_\mathrm{c}= 12 \,\mathrm{Myr} \sqrt{T/10^7\,\mathrm{K}}/ (n/\mathrm{cm^{-3}})$.
Hence, if the density is high enough to slow down the bow shock as required, the bow shock will be radiative, even if the atmosphere
has already keV temperatures. A thin and dense shell forms, that may provide the required column density.
Since the shell is subject to a number of instabilities, the shell fragments, and the radiation can be seen unabsorbed through the holes,
in the larger sources.

The natural column density of the shell, 
$N_\mathrm{sh} = 10^{23}$~cm$^{-2} (\rho/ 5 \,m_\mathrm{p}\, \mathrm{cm}^{-3}) (r_\mathrm{sh}/25\,\mathrm{kpc})$,
which 
\replace{}{
would correspond to a surprisingly large gas mass of order 
$10^{13}\,M_\odot$, and}
is much more than observed. However, the shell clumps quickly, and a clumpy multi-phase medium is also what is infered from 
detailed simulations of 2D radiative shocks \citep{SBD03}. In fact this may be the only way to explain that not only
atomic hydrogen and highly ionised carbon, but also molecular gas is observed at the same velocities \citep{dBrea03b,dBrea03a}.
The effective column will be reduced. The low column regions
might be ionised by the recently discovered inverse Compton (X-ray) emission of the cosmic microwave background from the radio cocoon,
which may well extend into the UV \citep{Scharfea03}.

Another issue is the Ly$\alpha$ luminosity of the shell, which should be considerably
below that of the emission line. Assuming one recombination per 
proton and a power law atmosphere $\rho=\rho_0 (r_\mathrm{sh}/r_0)^\kappa$, this is given by:
\begin{eqnarray}
L_{\mathrm{sh, Ly}\alpha}&=& 4 \pi r_\mathrm{sh}^2 v_\mathrm{sh}
h \nu_{\mathrm{Ly}\alpha} \rho / m_\mathrm{p} \nonumber \\
 & = & \left( \frac{h \nu_{\mathrm{Ly}\alpha}}{m_\mathrm{p}}\right)
\left[\frac{144 (\kappa+3)}{\pi^2 (\kappa+5)^2} L \rho_0^2 r_0^4 
\left(\frac{ r_\mathrm{sh}}{r_0} \right)^{4+2\kappa} \right]^{1/3} \nonumber \\
& \approx & 3 \times 10^{43} \,\mathrm{erg/s} 
 \left[ \frac{L}{10^{46} \,\mathrm{erg/s}} 
\left(\frac{\rho_0}{5 m_\mathrm{p} \,\mathrm{cm}^{-3}}\right)^2 \right. \nonumber \\ 
& & \left. \times \left(\frac{r_0}{25 \, \mathrm{kpc}}\right)^4
 \left(\frac{ r_\mathrm{sh}}{r_0} \right)^{4+2\kappa} \right]^{1/3}\label{lyarad}\enspace.
\end{eqnarray} 
The observational requirements may be fulfilled, postulating a high density inside of $\approx 10$~kpc, and a steep decline ($\kappa<-2$)
further out. Clearly, the atmospheres have to be carefully tuned. Recent estimates on jet powers easily reach $10^{47}$~erg/s
\citep{Ghi03,mypap02d}. Because high redshift radio galaxies are among the most powerful jet sources, this makes the model more difficult.

\section{Jet within a galactic wind's radiative bow shock}\label{jpw}
This model involves a galactic wind due to supernova activity in the host galaxy, probably related to a starburst. 
Due to the low power of such a wind compared to a typical jet power, it will only have an observational effect, if the wind starts long before
the jet activity. The wind would have a radiative bow shock, which is responsible for the aforementioned Ly$\alpha$~absorption in the case of 
small jets.
When the jets reach the shell, they destroy it so that the larger sources are no longer absorbed. 
The details of this process are studied in the simulation below. 

In such a model, the parameters can be well constrained from 
observations. For the postshock gas to have a shorter cooling time than the preshock gas, the Mach number should be below six.
This restricts the temperature of the atmosphere via the sound speed. A shock velocity of 200~km/s, implies an X-ray atmosphere
with $0.2-1.5 \times 10^6$~K, in agreement with the non-detection of thermal bremsstrahlung in 4C~41.17 \citep{Scharfea03}. 
The age of the galactic wind bubble would be 
$t_\mathrm{d}=3r_\mathrm{sh}/5v_\mathrm{sh}=
73\,\mathrm{Myr} \, (r_\mathrm{sh}/25\;\kpc) (200\;\mathrm{km/s}/v_\mathrm{sh})$.
This is to exceed the cooling time due to bremsstrahlung, and consequently the pre-wind density is limited to 
$n\gtrsim0.06\,\ccm \sqrt{T/10^6\,\mathrm{K}}\,  (25\,\kpc/r_\mathrm{sh}) (v_\mathrm{sh}/200\,\mathrm{km/s})$.
It could be even lower, if the cooling is to happen in the postshock gas only, and more realistic cooling functions are applied. 
Such a density is similar to local clusters of galaxies.
The resulting total absorbing column in the postshock gas would be two orders of magnitude below that of the previous model,
alleviating the above described problem, but still requiring the dominant fraction not to be HI.
The power required to drive the wind is:
$L> 5\times10^{43} \,\mathrm{erg/s}\, \sqrt{T/10^6\,\mathrm{K}}\, (v_\mathrm{sh}/200\,\mathrm{km/s})^4 (r_\mathrm{sh}/25\,\kpc)$,
equivalent to some supernovae per year. Starbursts in otherwise normal galaxies at high redshift show winds with similar parameters
\citep{Dawsea02,Petea02,Tapea04}.

\section{Simulation}\label{sim}
\subsection{Setup}
The jet inside a galactic wind scenario was simulated in 2.5D with the hydrodynamic code {\em NIRVANA} \citep{ZY97}. To the evolution equations for 
mass, momentum and energy, the force of a dark matter halo, and the zero metal,
zero field cooling function from \citet{SD93} was added.
\replace{}{
The number densities of electrons and ions, which determine the cooling rate,
are computed from the total density and temperature, 
in the approximation of a pure hydrogen plasma in collisional ionisation
equilibrium.}
The computational domain spanned $\left[Z \times R \right]=[200 \, \kpc \times100 \,\kpc]$, 
resolved by $[2046\times1022]$ cells. 
A control run was performed at double resolution, but spanned only 50~kpc
radially, which the first simulation showed to be sufficient.
The grid was initialised with an isothermal King atmosphere
($\rho = 0.3 \,m_\mathrm{p} \,\mathrm{cm}^{-3} / \left(1+r^2/100\,\kpc^2\right)$, $r^2=R^2+Z^2$) 
at a temperature of $10^6\,\mathrm{K}$. 
In order to break the symmetry, random fluctuations on the percent level have been added to the density.
The galactic wind was simulated by a distributed energy and mass density increase, $\propto \exp(-r/3\,\kpc)$, 
$3\times10^{43}$~erg/s, and $10\;M_\odot/\mathrm{yr}$ in total. 
With these parameters, both cold gas due to a cooling flow and a radiative bow shock by the galactic wind can be expected.
After 80 Myr, a bipolar jet with a radius of $1\;\kpc$, a density 
of $10^{-5}\;m_\mathrm{p}\;\ccm$, a Mach number of 13, a velocity of $2/3$~the speed of light, 
and a total power of $4\times10^{45}\;\mathrm{erg/s}$ was injected in the centre of the grid.
\begin{figure*}
\centering
\includegraphics[width=\textwidth]{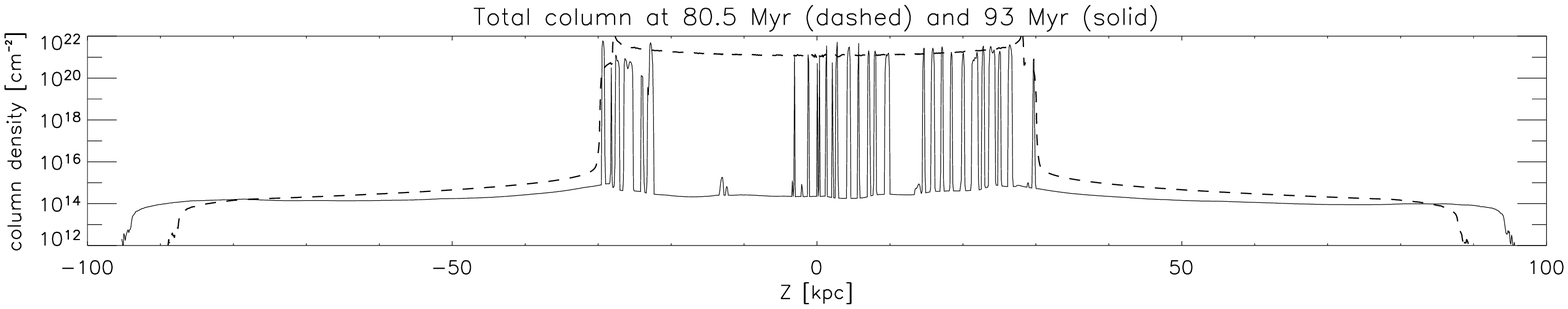}
\includegraphics[width=\textwidth]{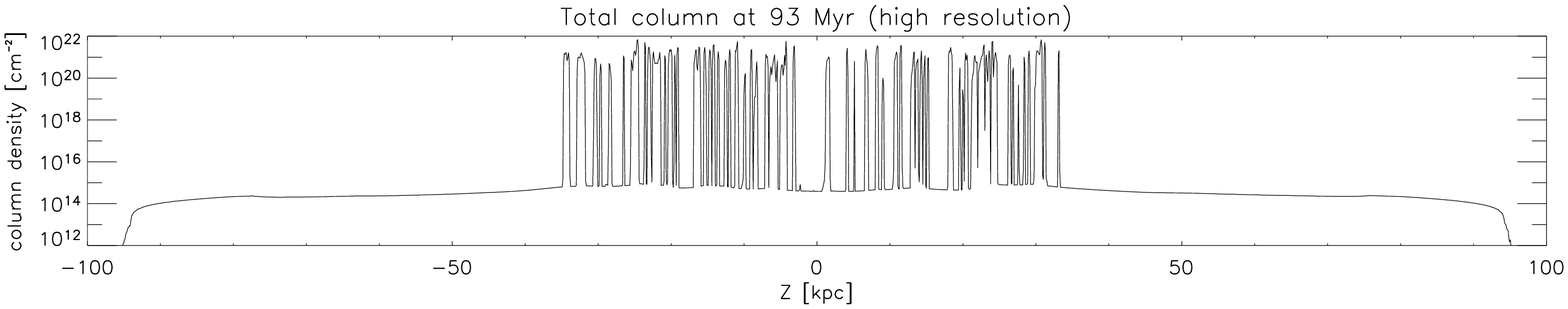}
\includegraphics[width=\textwidth]{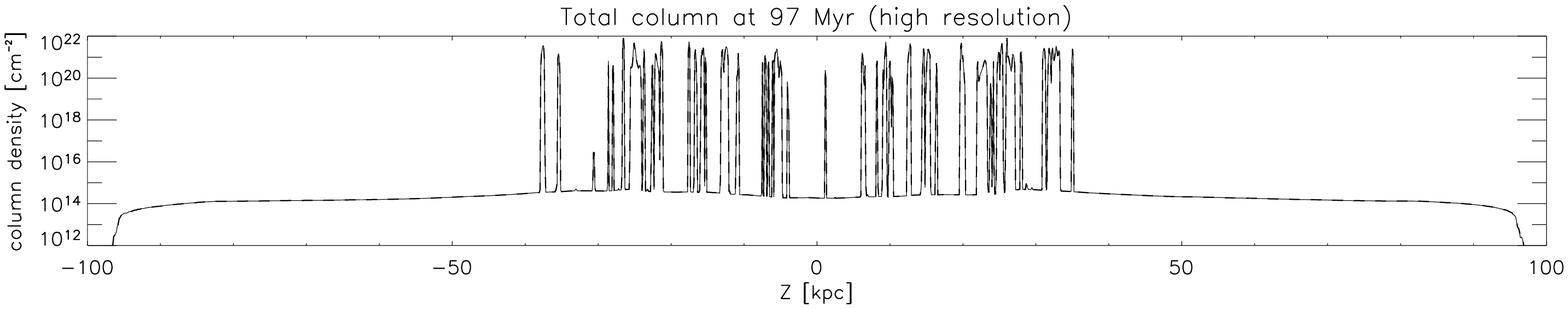}
\caption{\small Column over axial position 
(top: low resolution, dashed line: 80.5~Myr, solid line: 93~Myr;
middle: high resolution, 93~Myr;
bottom: high resolution, 97~Myr).
\label{cd2}}
\end{figure*}

\begin{figure*}
\centering
\includegraphics[width=0.49\textwidth]{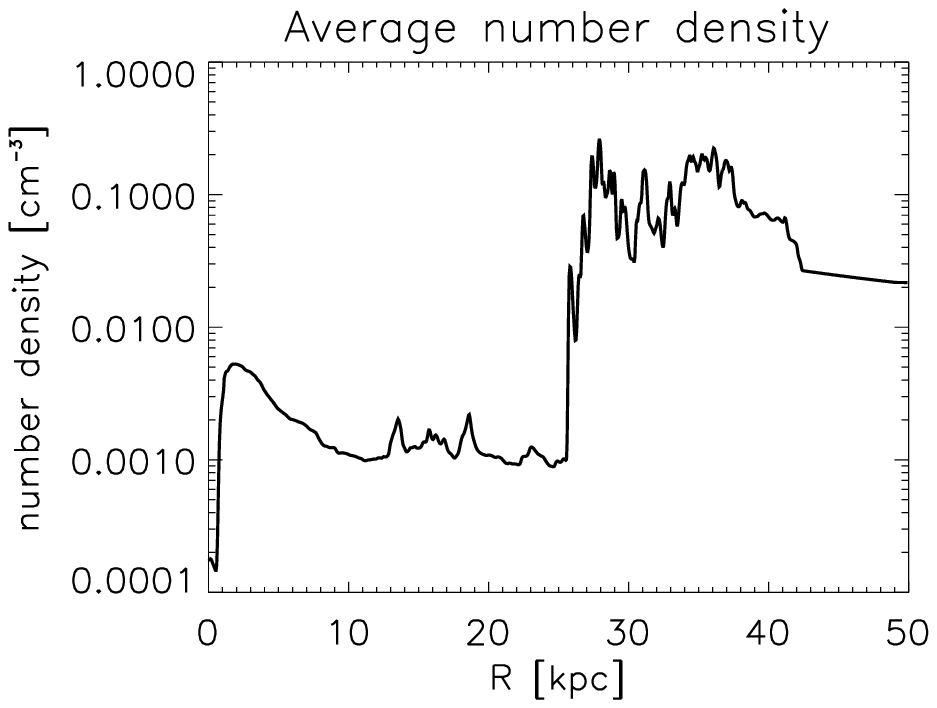}
\includegraphics[width=0.49\textwidth]{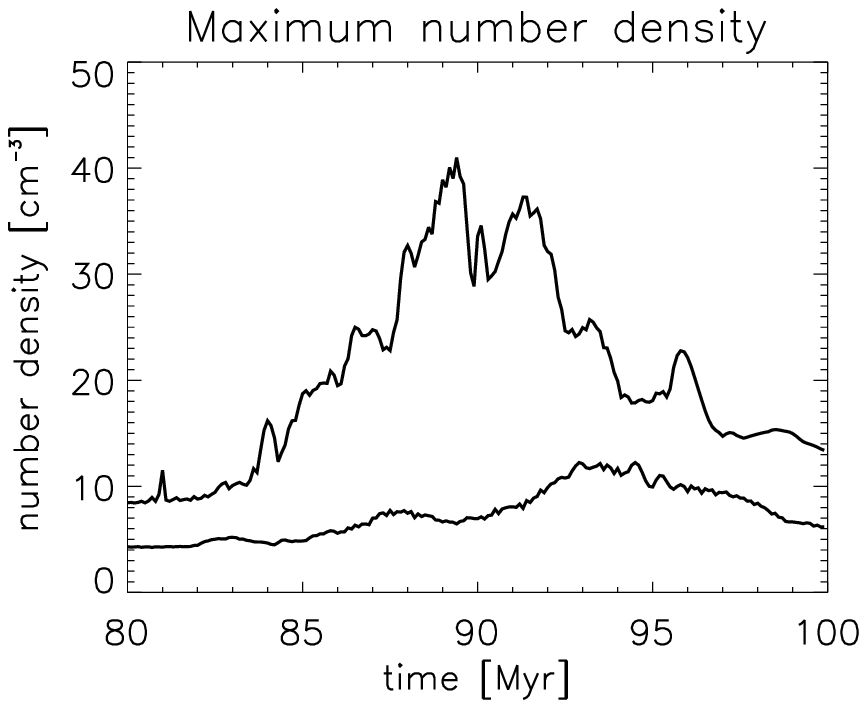}
\caption{\small Left: Average number density within the shell 
	(abs($Z$)$<28$~kpc) at 97~Myr for the high resolution simulation. 
	Right: Maximum density, always occuring in the shell, over time.
	The high resolution simulation takes the higher density values.
	\label{avn}}
\end{figure*}

\begin{figure*}
\centering
\includegraphics[width=0.49\textwidth]{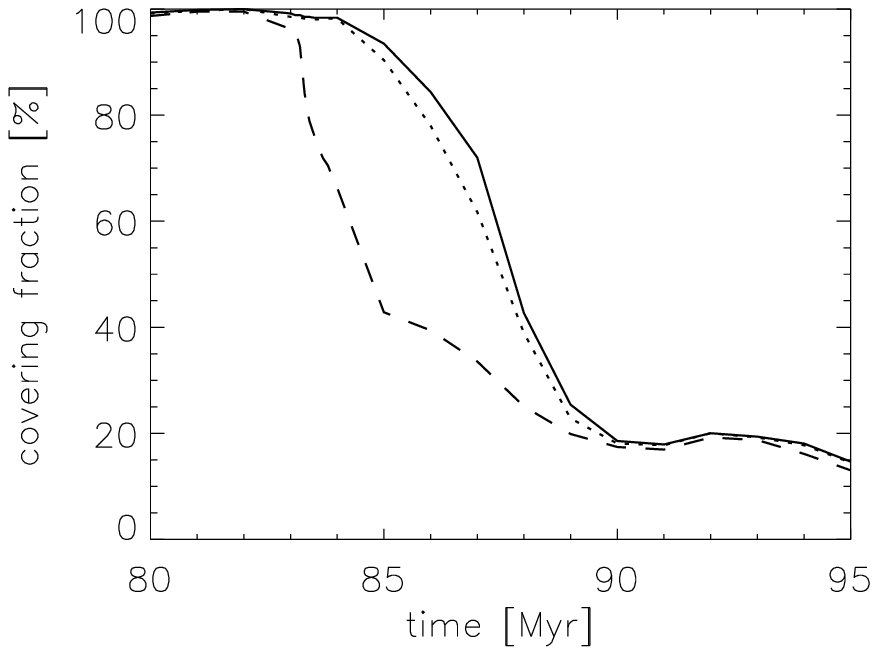}
\includegraphics[width=0.49\textwidth]{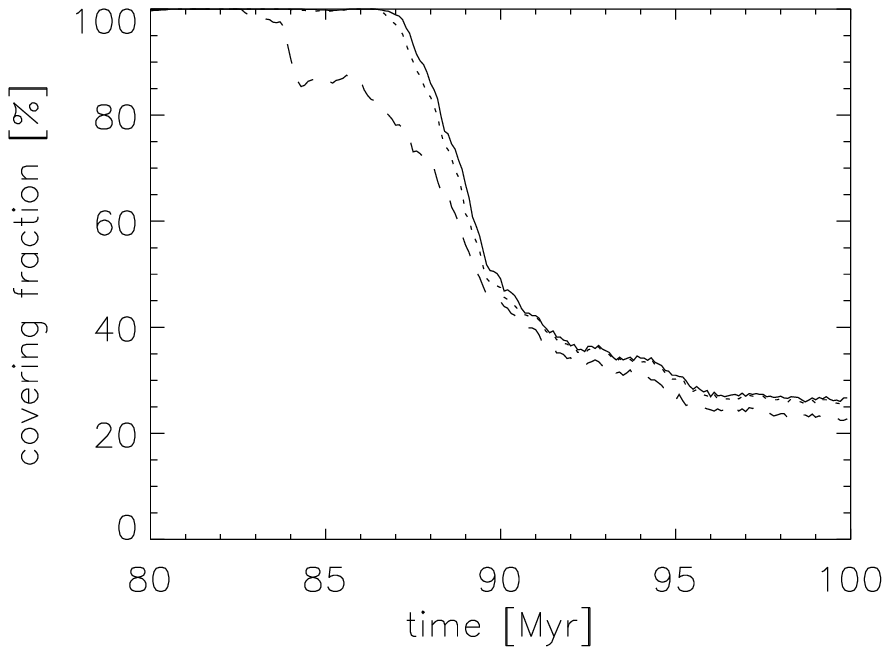}
\caption{\small Covering fraction of neutral hydrogen within $|Z| < 30$~kpc 
for a column in excess of $10^{18}$~cm$^{-2}$ (solid),  $10^{19}$~cm$^{-2}$ 
(dotted), and  $10^{20}$~cm$^{-2}$ (dashed). 
The left (right) panel shows the result for the low (high) 
resolution simulation. \label{cd3}}
\end{figure*}

\begin{figure*}
\centering
\includegraphics[width=0.99\textwidth]{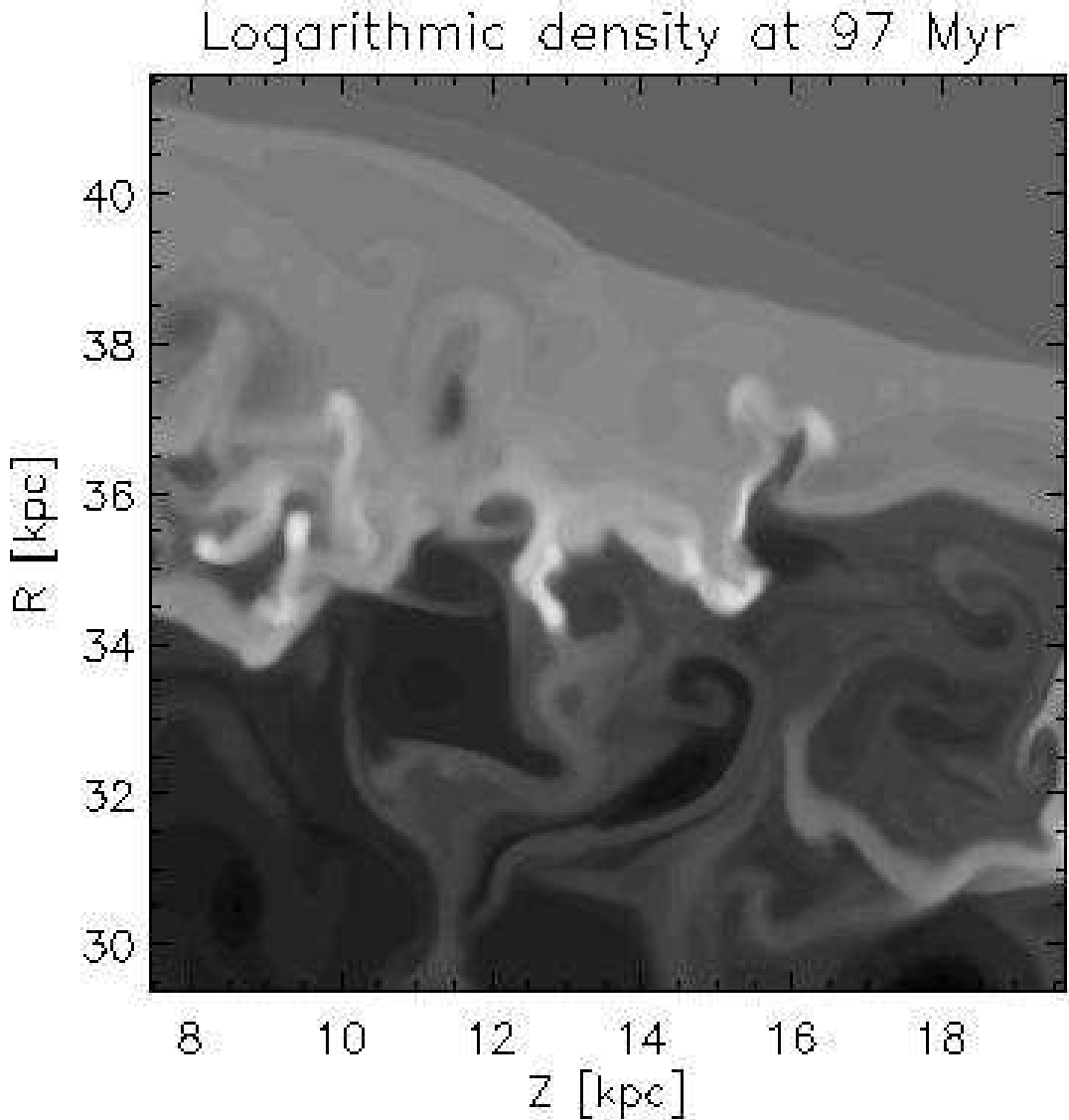}
\caption{\small Closeup of the shell fragments at 97~Myr 
	for the high resolution simulation. \label{closeup}}
\end{figure*}

\subsection{Results}
Density and temperature distributions at 80.5~Myr and 93~Myr are shown in 
Fig.~\ref{dtl} (low resolution simulation) and at 93~Myr and 97~Myr in 
Fig.~\ref{dth} (high resolution simulation).

During the first 80~Myr, the galactic wind establishes a radiative bow shock, 
which runs into a cooling flow atmosphere.
The radiative bow shock is marginally resolved, 
i.e. the temperature rises in the shock, and then cools to $\approx 10^4$~K,
where the cooling function was cut. Up to $\approx 81$~Myr, 
the jet stays inside of the wind shell. Then it quickly bores
holes at the heads, and the cocoon fills up the wind bubble. 
The increased pressure inside the wind bubble now accelerates
the shell, which is then destroyed due to Rayleigh-Taylor instabilities.

In this phase, large streams of gas, raining down from the shell into the 
interior of the bubble can be observed (Fig.~\ref{dth} and Fig.~\ref{closeup}).
Being stirred up by the jet's backflow, 
such gas may contribute to emission line gas. 
In the simulation, even at the higher resolution, 
this gas soon mixes with the very hot and rarefied plasma from the jet cocoon, 
but in reality it may stay unspoiled at emission line temperatures, 
contributing to the emission line halo.
Even so, the entrained gas condenses at the lower parts of the cocoon
next to the jet beam (Fig.~\ref{avn}), due to the background gravity.
The galactic wind also contributes somewhat to the high density in that region.
Also, the gas is constantly stirred up due to backflowing vortices.
\ignore{The two beams have different lengths, where the beam that is first shorter penetrates through the wind shell faster.
The armlength ratio is similar to other bipolar jet simulations \citep[$\approx 10$\%,][]{mypap04a}.}

The wind shell fragments into dense clumps. Hotter gas flows past these clumps and establishes a new, faster and non-radiative bow shock.
A closeup of the shell fragments can be seen in Fig.~\ref{closeup}.
The maximum density, which is reached in the shell, is shown over time in 
Fig.~\ref{avn}~(right). For the high resolution simulation, it starts to rise
at 81~Myr, when the tip of the jet hits the shell. Oscillating, it then keeps
rising as long as the shell remains intact. In that phase, the jet cocoon
compresses gas into the wind shell. At about 90~Myr, the fragmentation starts.
The cocoon material can then escape from the bubble, the pressure drops,
and the gas clouds expand again. The jet leaves the grid shortly after 97~Myr.
Data after that time could be affected by that. 
At lower resolution, the same things happen,
but take more time. At jet start and towards the end of the simulation,
the maximum density is about half the one in the highly resolved simulation.
This reflects the resolution increase by a factor of two. In between, the 
higher resolution simulation reaches a four times higher density than
the low resolution one. The simulation is obviously not numerically converged.
\replace{}{Nonetheless, even if not fully converged, the simulations
serve to illustrate the fundamental physical 
behaviour that is important in understanding the associated absorber 
phenomenon, namely that the jet-wind interaction causes the formation 
of many dense clouds through the operation of the R-T instability.
Furthermore, there is good reason to apply this kind of resolution,
which effectively cuts the R-T modes smaller than some critical wavelength,
as explained below.}

The column density distribution for neutral hydrogen 
has been analysed at 80.5~Myr and 93~Myr
(low resolution, Fig.~\ref{cd1a}) and 93~Myr and 97~Myr 
(high resolution, Fig.~\ref{cd1b}). 
The neutral fraction has been calculated according to collisional ionisation
equilibrium \citep{SD93}. 
The 2D representations (over Z-position and R-velocity) show that, 
as long as the jet is inside the wind bubble, both, 
the cooling flow (down to -100~km/s)
and the wind shell (200~km/s~--~400~km/s) are able to form significant
 absorption systems, totally covering the radio galaxy.
The infalling material has a very narrow velocity distribution. 
For the wind shell, the full width at half maximum is $\approx 100$~km/s. 
The highest column density
is found between 200~km/s
and 250~km/s. The speed of the radiative shock is 200~km/s.
At 93~Myr, the shell has been destroyed. At low resolution,
no significant column is left below
500~km/s, neither of the wind shell nor of the cooling flow. 
The surviving clumps still produce high column, but are not coherent,
neither in space
nor in velocity space.
At higher resolution, the clouds are more resilent. There is significant
column left between 200~km/s and 400~km/s, also at late times. 
However, this comes mainly from non-central regions, and at 97~Myr the central regions are devoid of slow neutral hydrogen.

At that time, the covering fraction is $\approx20\%$. 
It takes $\approx 1-3$~Myr for a significant drop 
(Fig.~\ref{cd2} and Fig.~\ref{cd3}).
At higher resolution, the covering fraction starts to drop later. It then drops
quickly to 50\%, 
after which it flattens, ending in a slightly higher number than the low 
resolution simulation.

\section{Discussion}\label{discu}
Using a galactic wind as the driver of the radiative shock reduces 
the lower limit on the external density 
to values observed in present day galaxy clusters. Also, the Ly$\alpha$
self luminosity (\ref{lyarad}) of the shell is low, 
as required for absorption systems.
The jet head pierced the wind shell after 1~Myr.
The simulation showed that the shell is able to totally cover the inner 
50~kpc of the jet with high column for 7~Myr after jet start.
At that time the jet is considerably more extended than the wind shell.
This means that the real wind shell would be smaller than assumed here,
since the radio jets are observed to have diameters of 50~kpc when 
the absorption disappears.

The jet impact accelerated the absorbing shell to a higher velocity region,
the shell thereby fragmented due to the Rayleigh-Taylor instability,
and part of the material fell into the interior
of the now radio cocoon filled shell. During this process the covering fraction
\replace{reduces}{reduced} to $\approx20$\%.
The simulations were 2D axisymmetric.
In reality, the reduction along the third dimension should be similar,
which would result in a final true covering fraction of 
about 20\%$^2 \approx 4$\%.
The emission lines could pass nearly unchanged through such a region.

\replace{}{
The assumption of axisymmetry also affects other parts of the simulation.
General considerations for jet simulations are valid also in this case
\citep{Norm93}, i.e. jets are less stable in 3D simulations, and magnetic 
fields as well as a high Mach number may be required for stable propagation
over the observed distances. An interesting extra is the interaction
of the cocoon with the wind shell fragments. This may well cause
asymmetries in the vortex shedding process, leading to the bending 
of the jet. This is observed in some of the larger sources
\citep[e.g.][]{vOea96,Pentea97,Pentea98}. Therefore, further investigation
is necessary.} 

The total neutral column is still higher than observed.
The reason may be that a significant fraction of the gas cools down 
even further to form molecular hydrogen
\citep[compare][ p338]{ADU}. 
When appropriately resolved, radiative shocks have recently been shown 
to become turbulent due to the thermal instability \citep{SBD03}.
Various ionisation species as well as molecular hydrogen might be present.
Details would require more involved shock models than presently available.

There are few examples of partially covered Ly$\alpha$ regions in 
the literature. 
\object{TN J1338-1942} might be a rare example \citep{dBrea99}.
The higher velocity clumps, present after shell destruction, 
may explain complex velocity profiles of larger sources. 

Doubling of the resolution raised the density
in the shell by more than
a factor of two. The shell fragments are more 
resilent at higher resolution \citep[compare][]{MKR02},
and therefore covering fraction and neutral column at low velocities
reduce more slowly.
The fragmentation of the shell starts earlier at higher resolution.
This is in agreement with the linear growth time for the 
Rayleigh-Taylor instability, which is proportional to the wavelength,
i.e. it starts faster if smaller wavelengths are resolved.
The simulation is therefore not converged. 
Convergence would be achieved by processes not included in the simulation.
The wind shell would be supported by its magnetic field,
limiting its density $n_\mathrm{s}$, and 
the Rayleigh-Taylor instability would finally be stabilised at the critical
wavelength $\lambda_\mathrm{c}$ by the transversal
magnetic field $B$ \citep{JNS95} of the radio cocoon:
\begin{equation}
\lambda_\mathrm{c}= 20\,\mathrm{pc}
\left(\frac{B}{10^{-5}\,\mathrm{G}}\right)^2
\left(\frac{10^{-7}\,\mathrm{cm/s}^2}{g}\right)
\left(\frac{10 \mathrm{cm}^{-3}}{n_\mathrm{s}}\right).
\end{equation}
Here, $g$ is a typical shell acceleration of 500~km/s in 10~Myr.
Depending on these parameters, the shell also may fragment faster or slower
in individual sources.
For these reasons, the applied resolutions seem to be reasonable. 

The simulation also shows that the blueshift of the absorber changes with 
position by $\approx 1$~km/s/kpc.
This is tiny, but might be detected in high resolution spectra. 
For sources near the critical size, high quality
radio observations at low frequencies may reveal the round central 
cocoon structure that is produced by the jet wind interaction.
Such observations may provide a test of the thin shell model.

\begin{acknowledgements}
This work was supported by the Deutsche Forschungsgemeinschaft (SFB 439).
The computations have been carried out at the
HLRS in Stuttgart (Germany).
\end{acknowledgements}

\bibliographystyle{aa}
\bibliography{/home/krause/texinput/references}

 \end{document}